\documentclass{article}

\usepackage{arxiv}

\usepackage[utf8]{inputenc} 
\usepackage[T1]{fontenc}    
\usepackage{hyperref}       
\usepackage{url}            
\usepackage{booktabs}       
\usepackage{amsfonts}       
\usepackage{nicefrac}       
\usepackage{microtype}      
\usepackage{lipsum}		
\usepackage{graphicx}
\usepackage{natbib}
\usepackage{doi}

\title{Interactive Storytelling for Children: A Case-study of Design and Development Considerations for Ethical Conversational AI}

\author{ {Jennifer Chubb}\thanks{Corresponding author} \\
	Department of Computer Science\\
	University of York, UK\\
	
	\texttt{jennifer.chubb@york.ac.uk} \\
	\And
	{Sondess Missaoui} \\
	Department of Computer Science\\
	University of York, UK\\
	\texttt{sondess.missaoui@york.ac.uk} \\
	\AND
	{Shauna Concannon} \\
	Centre for Research in the Arts, \\Social Sciences and Humanities\\
	University of Cambridge, UK\\
	\texttt{sjc299@cam.ac.uk} \\
	\And
	{Liam Maloney} \\
	Department of Theatre, Film, \\Television and Interactive Media\\
	University of York, UK\\
	\texttt{liam.maloney@york.ac.uk} \\
	 \And
	{James Alfred Walker} \\
	Department of Computer Science\\
	University of York, UK\\
	\texttt{james.walker@york.ac.uk} \\
 }

\renewcommand{\shorttitle}{\textit{CAI} for c
Children}

\hypersetup{
pdftitle={A template for the arxiv style},
pdfsubject={q-bio.NC, q-bio.QM},
pdfauthor={David S.~Hippocampus, Elias D.~Striatum},
pdfkeywords={First keyword, Second keyword, More},
}

\begin{document}
\maketitle

\begin{abstract}
Conversational Artificial Intelligence (CAI) systems and Intelligent Personal Assistants (IPA), such as Alexa, Cortana, Google Home and Siri are becoming ubiquitous in our lives, including those of children, the implications of which is receiving increased attention, specifically with respect to the effects of these systems on children's cognitive, social and linguistic development. Recent advances address the implications of CAI with respect to privacy, safety, security, and access. However, there is a need to connect and embed the ethical and technical aspects in the design. Using a case-study of a research and development project focused on the use of CAI in storytelling for children, this paper reflects on the social context within a specific case of technology development, as substantiated and supported by argumentation from within the literature. It describes the decision making process behind the recommendations made on this case for their adoption in the creative industries. Further research that engages with developers and stakeholders in the ethics of storytelling through CAI is highlighted as a matter of urgency.
\end{abstract}


\section{Introduction}

Conversational AI (CAI) agents are ubiquitous in the lives of adults and children across the developed world. Intelligent Personal Assistants (IPA) such as Cortana (Microsoft), Alexa (Amazon), Siri (Apple), and Google Assistant are perhaps the most well known form of CAI and are at the forefront of technological advancement. CAI has become more effective thanks to advances in automatic speech recognition (ASR) \cite{karpagavalli2016review}, Natural Language Processing (NLP) \cite{trilla2009natural,vanzo2019hierarchical}, and Deep Learning (DL) models \cite{abdel2014convolutional}. The fast paced evolution of Artificial Intelligence (AI) has led to the regular use of high performance CAI systems in day-to-day activities. CAI software enables individuals to communicate with a wide range of applications in natural language via voice, text and video. Researchers have begun to explore how these technologies are embedded within family practices and how interactions differ when involving adults and children (e.g. \cite{sciuto2018hey,druga2017hey}).
Children start engaging with the internet and technology at a very young age for entertainment, education and social reasons. For instance “already one in four children between 5 and 16 years of age live in a household with a voice-activated virtual assistant in the UK” \cite{childwise2019}. 

However, for younger users this is often without necessarily being aware of the associated risks \cite{livingstone2011risks,long2020ai}. Sadly, children can easily access inappropriate content, or be manipulated online through communication technologies \cite{dcms2019harms}. From a young age, children are learning what it means to develop and build relationships, establishing their place in the world. The nature and role of that interaction and the ultimate relationship shared between children and CAI agents demands attention. Children’s uniqueness is especially pronounced  when we consider the stages of a child’s development and their interaction with technology. It is therefore extremely important to account for such differences. Issues such as confidentiality, representation, bias, responsibility, trust and veracity, power and freedom related to CAI therefore become especially pertinent.

\subsection{Defining AI in the context of conversational agents}

AI is often referred to as an ‘umbrella term’ encompassing a range of tools inclusive of Machine Learning (ML), NLP and DL. Advances in AI have opened up the possibility of developing new forms of engagement, e.g. news, storytelling and interactive forms of entertainment \cite{stefnisson2018mimisbrunnur,riedl2011game,thorne2020hey}. While ethics increasingly dominates the AI literature, specific considerations of interaction design that ensures the safety of children provokes the need for more urgent ethical reflection \cite{frauenberger2019broadening}.
Our case-study addresses CAI challenges with respect to privacy, safety, security \cite{mcreynolds2017toys}, and the effects on education and health domains \cite{de2020intelligent}. For the purposes of this paper, we use IPA to refer to voice enabled personal assistants, and CAI to collectively refer to all systems that facilitate interaction in natural language (e.g. text-based chatbots). 

To the best of our knowledge, few studies have combined considerations of ML, NLP and DL innovation for CAI with a mapping of the ethical implications presented in the literature in the creative industries. Using a pilot case-study, we describe and reflect on the ethical design of a CAI meta-story  tool for children’s storytelling. By exploring previous research on both technical and ethical aspects, this paper reflects on the design and development decisions we made supported by argumentation in the literature. In doing so, we propose deeper and richer analysis of the issues for children’s storytelling CAI in the creative industries.

This paper begins with an overview of the ethical issues currently discussed with respect to children, both in policy and academic literature. This is then related to a mapping of the technical advances in the general area of CAI, focusing on acoustic models and data-driven models and the ethical considerations thereof as applied to our case-study - the development of a meta-story chat tool.

\subsection{Emerging immersive AI technologies in the creative sector}

Creative industry practitioners are looking to develop innovative and engaging experiences for children. As new forms of storytelling and immersive experience emerge, and virtual, mixed, diminished and extended reality projects become more commonplace the need to examine the associated risks becomes more pressing. Children may be encountering these technologies while they are still forming how they discern the difference between reality and fantasy (e.g. the use of Sesame Street in Stanford University’s Virtual Human Interaction Lab, Virtual Reality 101)\footnote{ https://www.commonsensemedia.org/research/virtual-reality-101}. While certain aspects of the creative sector such as the ethics of games and children is relatively well researched \cite{cano2015agile} including a range of work on parental concerns and consent \cite{dixon2010parents,willett2015discursive,rode2009digital} including their gamified uses even to teach ethics \cite{bagus2021designing}, what happens with respect to children’s data as they interact with voice technologies for entertainment, poses deep moral concerns. Recent work suggests that such immersive experiences reveal a range of social issues including social isolation, desensitization, depersonalisation, manipulation, privacy and data concerns \cite{bailey2017considering,grizzard2017repeated}. The more widespread these immersive storytelling tools become, the greater need there is to reflect deeply on their design, in particular for children. Long and Magerko \cite{long2020ai} highlight the importance of AI literacy, i.e. the competencies that enable individuals to critically evaluate and collaborate with AI technologies, and demonstrate the variety of factors that influence children’s perceptions of AI. This is critical to the ethical design of CAI and a crucial aspect of child-computer interaction.  Indeed, there is a need to empower children in the design process through participatory approaches relevant to the child-computer interaction field \cite{kumar2018co,yip2019laughing,piccolo2021chatbots}.

For creative sector organisations, many of which are SMEs, simultaneously directing attention towards the development of exciting and engaging experiences and ensuring the ethical and safe deployment for children (which as highlighted poses a number of unique considerations), can be a daunting endeavour. Furthermore, the over-abundance of ethical guidance documents, coupled with the limited mapping of these high level principles onto practical implementation strategies makes this a difficult space to navigate, especially with respect to children. Researchers have highlighted how ethical guidelines often fail to acknowledge the important practical difficulties of implementing AI systems or the additional work required to translate these high level principles and their various implications into actual workflows \cite{ryan2020artificial,tomalin2021practical}. AI in the creative industries and digital storytelling in its current manifestation presents, at best, an inconsistent approach to responsible innovation of CAI for children, often with a need to join up the ramifications of situating such technologies within the home with the consequential impacts on users (children). 

The inherent biases and assumptions underpinning current technical methodologies require the utmost scrutiny when applied to vulnerable groups such as children.  As storytelling is a universal way of connecting with others and in the case of young people, these connections are vital to their mental wellbeing, safety, education and enjoyment. 

\subsection{Momentum in AI ethics}

Responsible innovation in science and technology has a long history \cite{bush2020science} but it is also a current issue and one with a newer research focus \cite{owen2013responsible}. There is also a growing interest in bridging the gap between AI practice and governance \cite{bryson2020artificial}. This is reflected in the publication of a significant number of ethical guidance documents emerging from both commercial and academic sectors \cite{morley2019overview,hagendorff2020ethics}. The global political landscape also attends to issues concerning ethical AI e.g. see the European Commission's White Paper on AI \cite{ECWhitePap} and the Children’s Online Privacy Protection Act (COPPA) in the US.

Perhaps the most active in the policy area of online harms and children is UNICEF (2020) and UNESCO the latter of which, embarked on the development of a global legal document on the ethics of AI for children (2021)\footnote{Elaboration of a Recommendation on the ethics of artificial intelligence}. The recommendations made by UNICEF include the need to closely examine privacy, safety and security by providing identity protection, detecting harmful content, focus on location detection and biological/psychological safety. Additionally, UNICEF is clear that inclusion and equitability are upheld - ensuring that systems are checked to mitigate against historic bias which may stand in the way of children’s fair chances in life. In this respect, biases might include health, education, credit, financial status of family etc. Dignity should be upheld with respect to automation of roles in the future and finally, the cognitive and psychological implications of technology with respect to mental health and manipulation should be explored. They suggest that a range of actors across the AI community including scholars and agencies, need to come together to engage with these concerns. The UK Centre for Data Ethics and Innovation, called for participatory design of smart speakers and voice assistants stating that ‘[u]sers are expected to be active participants in the development of these technologies’ \cite{Snapshot-CDEI}. They suggested that users should actively ask questions of their devices about how their data is used and stored, and even exert market influence to drive up demand for privacy preserving technologies. However, participatory approaches in ethical design which actively consult stakeholders, children and young people is a positive and progressive approach \cite{cortesi2020youth,kumar2018co}.

We draw on argumentation from the academic and policy literature, to describe four emergent themes which guided the development and design of a meta-story chat tool for children. The themes which guided the co-production of this tool include: to consider the effects of CAI on the cognitive and linguistic development of children; moral care; inclusivity; and regulation.  This paper aims to provide a lens through which to consider broader and deeper considerations for the responsible development of CAI for children’s storytelling. Seeded by our work with this pilot study, we aim to highlight several themes with accompanying discussion that inform the development of responsible CAI and to promote thought on future research. The following sections present the findings of the technical and ethical scoping work.

\section{Methods}

\subsection{A pilot case-study: AI Fan Along}

The focus of this paper is CAI for children’s storytelling and it reflects on a research and development pilot project to design a meta-story chat tool. We present a pilot case-study of work conducted with a digital agency committed to the responsible innovation of child-friendly CAI technology called ‘AI Fan Along’. The project was motivated by asking what the guiding ethical questions and principles pertinent to the design and development of CAI for children are and how they map onto its innovation. In order to investigate and answer these questions, we undertook a pilot-study involving background research to understand the most recent developments in the design and development of CAI for children from both technical and social perspectives. This led to the recommendations mapped out in the paper.

\subsubsection{Designing a meta-story tool for children}

The case-study which is the subject of this paper refers to the prototype ‘AI Fan Along’ - a meta-story chat tool to encourage children (ages 9-14) to engage with characters, storylines and issues using voice AI technology. The overarching aim of the platform was to increase social development within children, focusing on developing higher levels of social, literary and empathetic understanding through immersive digital storytelling. The tool would allow children to safely engage with their favourite characters on TV shows through voice-assisted technology and was designed so that when an episode of a TV programme ends, a child will be encouraged to speak to the characters to reflect on the events and participate with suggestions and predictions for the next episode thereby directing the narrative. To place children at the heart of the storytelling experience in an immersive way through voice technology was acknowledged by the developers as potentially harmful, raising a number of ethical considerations such as consent and privacy. Through research and development, the research team worked together to co-develop the technical design and ethical aspects of this prototype. In the following, we explain the process and methodology that was adopted to develop these recommendations. This pilot project was carried out in 2020, over a three month duration with academic and industry partners. The research was funded during the time of a national lockdown in the UK due to the COVID-19 pandemic. Our approach was two-fold; to conduct research on the technical potential of the tool and research on the ethical implications of these technologies for practice. 

\subsubsection{Conducting research into ethics of CAI}

From the perspective of ensuring ethical design of the tool, and in order to get a richness of perspectives on the effects of the tool, the team’s original research plan involved interviews with children and their parents testing the tool and the analysis of transcripts. Due to the pandemic, the design had to be adjusted and gathering qualitative data was not possible. Instead, the methodology was adapted to include research on the ethics of CAI for children. This included a non-exhaustive but thorough review of the current literature which resulted, through thematic analysis \cite{clarke2014thematic}, in guiding themes which aided the development of principles and ethical reflection for both the company and the researchers.  

Keywords developed to guide the non-exhaustive mapping of the literature on the ethics of CAI for children from recent years (up to 5), concurrent with a review of research on the technological research advances in CAI different categories included: CAI, ethical implications/ethics, children, young people, generations, safeguarding, impact, ASR, systems for conversational speech, voice assistants, Alexa, Google Home, Nest, Chat. The research team met through regular meetings which resulted over the three month period in two working papers covering both the technical and ethical aspects of the work. The ongoing iteration of the findings throughout characterise this case-study as a co-production project, whereby there was ongoing dialogue and ethical reflection between the research and development team.

\subsubsection{Limitations}

It is important to note the limitations of this research and the associated approaches. We aimed to devise a set of recommendations for the industry partner in a very limited time-frame. We do not have user experiences as a result of the adjustment to our methods within the given time-frame and acknowledge that further research will deepen our understanding by engaging with children and their parents through ethnographic or semi-structured interviews. The searching of the literature, though thorough, was not fully exhaustive or systematic in nature, again owing to the time and scope of this limited pilot study. As such, findings from this project may be limited in their generalisability. We aim to show how investigations of other technologies informed our design. We explore this by examining the technological options in CAI design supported by the literature. 

\section{Exploring technological options in CAI design}

Even though CAI could be an effective tool to aid children in their cognitive, social, and linguistic development, their didactic potential in storytelling context is not well investigated. The effectiveness of voice assistants in storytelling for children could be highly influenced by technical implementation of the chosen technology. In working on this case-study project it was necessary to review the technical implementations of CAI, as different methods pose distinct ethical challenges and the forms of interaction the system aims to support would require different architectures (e.g. answering questions about a specific book or TV show, through to more open ended forms of dialogue). For instance, to develop a customized meta-story  tool, which would engage children with their favourite TV show, we found it was important to consider children's linguistic development challenges. In particular, ‘AI Fan Along’ needed to support the child's ability to express and understand feelings through an adapted technology.  

A mapping of ML, NLP and DL innovation in CAI technology and the implications for the design and deployment of voice cloning systems for children was undertaken including a review of the most popular tools and frameworks in use by both industry and academia. This included research of current practices and ongoing co-production with the industry partner. Similarly, research on the audio aspects of the tools development was conducted, with a particular focus on ASR systems and their compatibility with child voices and physiology, and the viability of voice cloning technologies to allow diegetic immersion to be maintained. Regular meetings ensured good dialogue and knowledge exchange at all stages.
 
We mapped the literature in audio and speech using keywords: voice cloning, voice modelling, speech synthesis, deep fakes and voice spoofing, and performed searches concerning  AI innovation using keywords; CAI, ASR, ML and voice assistants, neural approaches to conversational AI; DL models; NLP and IPA. We first describe the background to this work before describing the design choices. 

\subsection{Outlining CAI technologies}

We aimed to provide our industry partners with a full picture on the CAI architecture and existing advances that could be easily adapted for AI Fan Along. 
As CAI requires the coordination and integration of several discrete systems performing pseudo-simultaneous tasks, we started by depicting the high-level architecture of CAI (see, Fig 1), as it is important to understand the potential role of each of them on the direct interaction between children and the voice assistant. Typically, CAI systems include an Automatic Speech Recognition (ASR), Natural Language Understanding (NLU), Dialogue Management (DM), Natural Language Generation (NLG), and Text to Speech (TTS) modules, which together constitute the high-level architecture of CAI. 

\begin{figure}[tb]
  \begin{center}
    \includegraphics[width=0.99\textwidth]{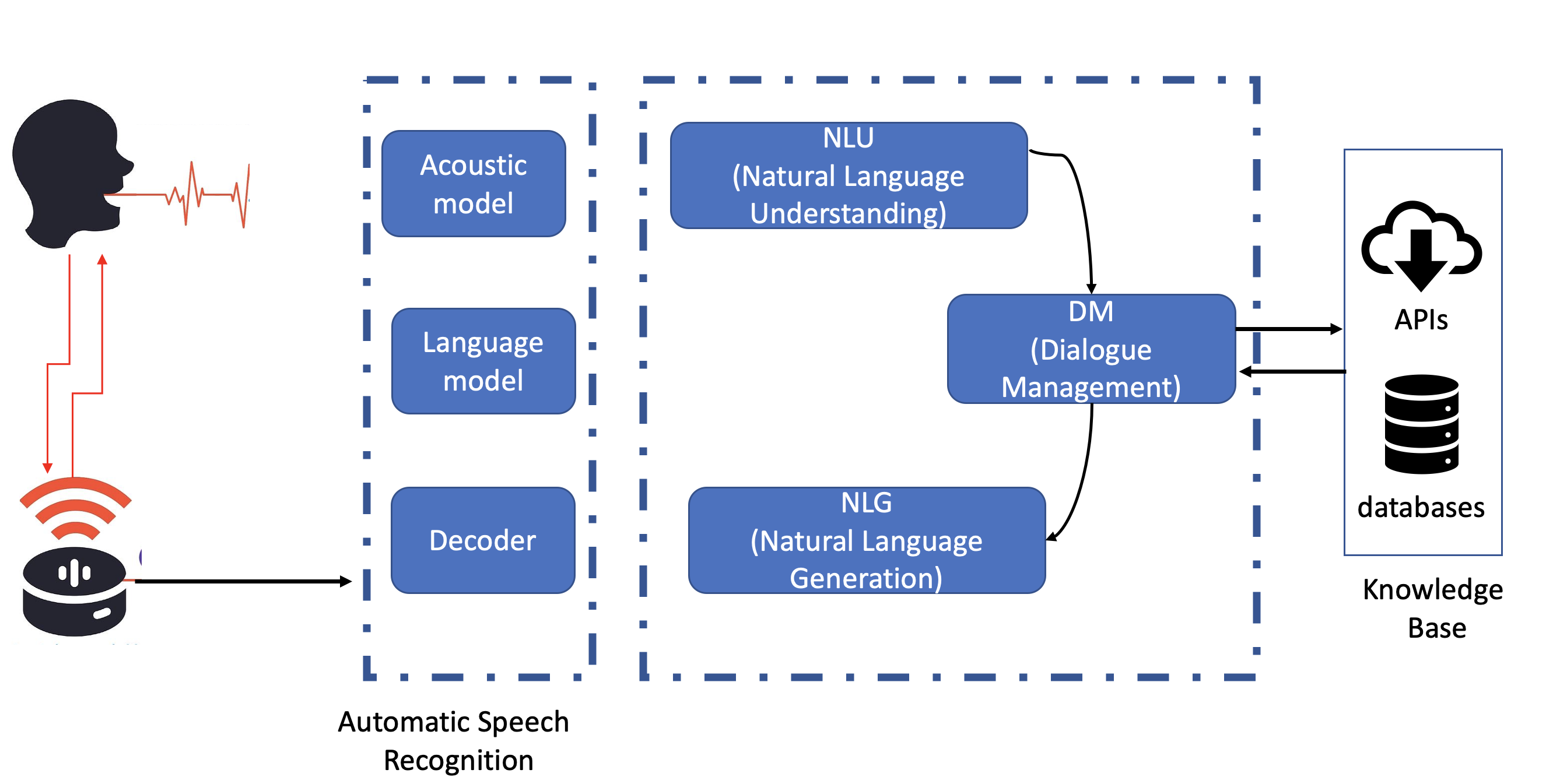}
  \end{center}
  \caption{A high level architecture for voice-based CAI}
\label{archi1}
\end{figure}

It is important to highlight that the NLU, DM and NLG components collectively comprise the semantic layer and are responsible for inferring meaning from the input, determining an appropriate next action and generating meaningful responses to output in natural language. Design decisions are typically informed by the type of interaction the system seeks to support. This was particularly the case with our case-study as the CAI needed to support task-based and open components. Dialogues are typically classified as task-oriented, i.e. supporting the user in completing a specific task, or open-domain, i.e. able to speak on a range of topics as determined by the user. Different implementations invariably require distinct considerations, may be suited to support different types of dialogue, and pose unique challenges.

\subsubsection{Task-oriented dialogues Vs open-ended dialogues: how to make the choice?}

CAI applications for children may encompass task-oriented and/or open-ended dialogues to support functional, educational, or entertainment-related interactions. However, selecting from the wide range of existing approaches in the case of ‘AI Fan Along’ was motivated by ethical concerns. In the following, we explore approaches used for implementing task-oriented and open-ended dialogue systems to identify their potential adaptation for AI Fan Along.
 
The NLU is a core component that interprets the meaning that the user communicates and classifies it into proper intent \cite{li2017investigation}. Rule-based approaches \cite[e.g.]{yaman2008integrative,schapire2000boostexter} have been widely used for both classifying the user’s intent and defining the system’s action, i.e. what is said. Rule-based approaches often follow an established set of dialogue-flows or handcrafted rules. This enables the system to respond effectively to a specific domain (i.e., task-oriented dialogues), but may be less effective if users pose questions.  Frame-based approaches use a template model to offer a more flexible approach. Consequently, the dialogue flow is not pre-determined, but adapts and incorporates the user’s input, and can integrate additional information sources from either the dialogue history or an external database. For example, Question Answering (QA) systems draw on techniques from Information Retrieval (IR) to enable the user to receive a relevant answer to a question asked in natural language, with sufficient context to validate the answer \cite{hirschman2001natural}. QA agents employ large-scale Knowledge Bases (KB) or a document collection in natural language to retrieve information that then populates ‘slots’ in the dialogue, to provide concise and externally validated answers.

QA systems have been employed for public engagement and entertainment purposes in culture and heritage contexts (e.g. \cite{robinson2008would}), and effectively enable users to navigate the KB through conversational interaction. Such frame-based approaches have also been used in open-domain dialogue contexts, such as the ALICE chat-bot developed using AIML \cite{abushawar2015alice}.

While designing a task-oriented dialogue system to assist users in performing a specific task (e.g., making a hotel reservation) requires a relatively constrained set of conversational possibilities, as this topic scope increases so will a system’s complexity. A drawback of these approaches is that they have limited adaptability and a challenge can arise when user utterances fall beyond the scope of the dialogue-flow or domain of expertise (i.e., the used KB). Additionally, even when the scope of an agent is clearly communicated, users often persist in confronting them with off-topic or `out-of-domain’ talk \cite{robinson2008would,ameixa2014luke}. 

In the case of AI Fan Along, a child’s speech behaviour is more variable than adults. While adults have been observed to modify their speech when interacting with CAI, e.g. using shorter and simpler phrases \cite{mou2017media}, the same can not be assumed for children. It is highly expected that children will produce unformulated and  unthought-out requests to ‘AI Fan Along’ as if it is a human. Approaches for managing off-topic talk include changing the topic and integrating a retrieval component using additional responses drawn from a corpus of film dialogues \cite{ameixa2014luke} could be particularly important to the design of a meta-story  tool. The literature reveals several attempts to understand child civility with machines and spoken dialogue systems \cite{burrows1992not,potamianos1998spoken,arunachalam2001politeness}.

On the other hand, recent advances in Deep Learning (DL) and the availability of large conversational datasets have made open-domain dialogue systems, capable of generating content on a wide range of topics, more viable.
Open-domain dialogue systems rely mainly on data-driven models and end-to-end (E2E) approaches \cite{roemmele2011choice,sutskever2014sequence,vinyals2015neural}. These have seen great success due to the availability of benchmarks (e.g., ConvAI Competition\footnote{https://convai.io/}), and pre-trained language models such as BERT \cite{devlin2018bert}. One of the advantages of data-driven models is the lack of dependencies on external resources such as API calls or KB. Moreover, these models can be totally trained from scratch independently from the NLU, DM and NLG components, which often require extensive domain expertise and contain limited design choices. Consequently, E2E systems demonstrate great promise for generating conversation on a more diverse range of topics as they require less sophisticated annotation schema.
Overall, data-driven models can be more flexible than rule-based systems, which make them more suitable for engaging in open-domain and social dialogues.
 
We found that although fully data-driven models are promising, they pose several challenges - particularly noteworthy with respect to our use-case. Neural response generation has a high likelihood of generating uninformative responses, e.g. “I’m not sure I understand”. According to \cite{li2016diversity}  this issue is due to the training objective or a bias that emerges from the training data itself \cite{li2016diversity,serban2016hierarchical,zhang2018generating}. Efforts to develop E2E systems capable of generating more naturalistic responses have included the development of datasets addressing more social and human-like aspects of dialogue. This was important to investigate for the use-case of AI Fan Along. 
We found that the use of personas, as in Zhang et al. \cite{zhang2018personalizing} and Lin et al. \cite{lin2020caire}, could be a suitable solution. However, ensuring appropriate responses are generated consistently remains a challenge. For instance, Lin et al. \cite{lin2020caire} point out that these approaches can still result in the development of morally dubious agents, who do not “have any sense of ethical value due to the lack of training data informing of inappropriate behavior” \cite{lin2020caire}.
By reviewing the state-of-the-art work in CAI design, we were able to highlight the potential of using a hybrid approach, using data-driven models that are tailored to specific personas together with rule-based approaches, which would need to be iteratively tested for safety. This would enable us to design a system that could respond safely and flexibly to children's conversational patterns and adequately parses out-of-domain talk.
This choice led us to investigate other important challenges in the field, namely how can a child-friendly CAI relate to childrens’ specific speech patterns. Therefore, we investigated the role of the ASR, voice synthesis and voice cloning techniques with a view to  enhancing the effectiveness of the chat tool.

\subsection{ASR, voice synthesis and cloning, and human-centred variables}

One of the most distinctive aspects of ‘AI Fan Along’ was its acoustic features that would enable it to maintain comprehensive and engaging conversation/interaction with children. The literature highlighted the importance of developing a tool that accounts for and understands the highly varied inconsistencies and mutability of children’s language. Hence, AI Fan Along required an ASR module built to intentionally learn from the ways children speak. The following goes deeper into features of ASR and voice cloning to distinguish possible challenges to be considered for adaptation of existing technology in our context.

Automatic speech recognition is a core element in CAI that has a direct impact on the quality of interaction. ASR is the process that translates user-spoken utterances into text. The performance of an ASR system depends mainly on the robustness of its components, however, its ability to successfully handle the variability in the audio signals play a key criterion. Here we outline the ways in which many CAI designs and systems are more appropriate for adults and do not fully consider the physical and physiological development of children in their design.
ASR faces several sources of acoustic variability \cite{yu2016automatic}, which is caused by complicated interaction and speaker characteristics. These can be categorized as: firstly, \textit{within speaker variables}, these concern momentary and longitudinal variations in the voice due to emotional expression and arousal \cite{Lee2004}, illness, age \cite{Vipperla2010,Morris1994}, body mass \cite{Souza2018} etc. All these factors need to be accounted for by the acoustic model to be representative of all potential speakers in all states. Secondly, \textit{between speaker variables} (i.e. variations in spoken language, vocal tone and speech style) which mainly concern different accents, non-native accents, dialects, slang, speech impairment and disorders, gender \cite{Swartz1992,Morris1994} and even race\cite{Xue2000}. The issue of speech impairment is particularly relevant in the case of children whose speech and articulation are still developing. Usually, children over-enunciate words, elongate certain syllables, punctuate inconsistently or skip some words entirely. Their speech patterns are not beholden to the patterns used for training systems built for adult users. Collectively, these variables impose a significant logistical challenge and necessitate substantially broad training data to provide any sense of accuracy.

Moreover, the \textit{audio quality} factor (i.e. the quality and clarity of speech received by the ASR device) also creates a possible technological bottleneck. The positioning of microphones within a physical CAI interface/device and the qualities of the space in which a device is placed (in addition to the position of the device within the space) are a critical factor that can influence the intelligibility of speech. The microphone directivity (polar pattern), arrangements of multiple directional microphones in an array, and frequency response(s) of said microphones employed within the device may necessitate different post-processing to any received speech, as will the method of transduction (dynamic, electret, or boundary-design) \cite{borwick1990microphones}. Furthermore, the relative distance factors, critical distances, and the reverb time (RT60) and average absorption of the space will impact the intelligibility of any received speech. Finally, the shape of surrounding material, absorption coefficients of surrounding materials, and environmental noise within the space present another potential hurdle for ASR systems. Simply expressed: placing a CAI device on a high countertop in a reflective space such as a kitchen may preclude children from interacting with the system simply because of acoustic features and transduction methodologies.

Within many CAI agent interactions a spoken response from the agent to the user is often required e.g. responding to questions, observing reminders, timing information. To generate these responses, several common systems are employed. Voice banking and phrase banking have been in use in various systems, notably in telephony systems and for individuals with vocal disabilities \cite{veaux2013towards}, for several decades. However, the systems have been superseded by synthesis approaches that produce naturalistic intonation and rhythm patterns. These systems can be divided into Text-To-Speech (TTS) that generate the text-based semantic content of the phrases spoken, and the synthesis components that generate the corresponding audio i.e. the ‘spoken text’.

The TTS synthesis procedure and acoustic models are major elements of ASR and any improvement towards CAI for children needs to consider them. In particular the TTS is a sequential process that produces a speech utterance from an input text involving a set of high-level modules \cite{reichel2006text}.
A lot of advances have been achieved in TTS development including WaveNet \cite{oord2016wavenet}, TACOTRON \cite{wang2017tacotron}, and Deepvoice. For instance, researchers from Baidu’s Silicon Valley Artificial Intelligence Lab have presented three iterations of their Neural TTS system \textit{DeepVoice} \textit{1, 2} and \textit{3} \cite{arik2017deep,gibiansky2017deep,ping2017deep}. All three system iterations share a core architecture based on a segmentation model for locating phoneme boundaries with deep neural networks, a grapheme-to-phoneme conversion model, a phoneme duration prediction model, a frequency prediction model, and an audio synthesis model using a variant of WaveNet. Constructed entirely from deep neural networks, DeepVoice 2 allows synthesis of speech from multiple speakers, with a significant improvement in audio quality. More interesting systems have been proposed taking the DeepVoice system as baseline, these include the neural voice cloning system \cite{arik2018neural} by Baidu research lab, representing a big step toward personalized speech interfaces. It learns to synthesize a person’s voice from only short fragments of audio by applying speaker adaptation and speaker encoding approaches \cite{karpagavalli2016review,trilla2009natural}. 
Besides achievements in the TTS and ASR field, existing systems are not designed for use with children, whose voices, and speech behaviour are more complex than that of adult users. The ability to replicate or otherwise synthesize a range of possible respondents in a CAI system raises important questions and challenges concerning inherent bias, race, gender, disability, nationality etc. These questions, arguably some of the most pressing considerations when working with children and CAI, are explored in greater depth in the discussion. 

We now draw together the themes noted in the academic and policy literature with respect to ethical design of CAI for children and discuss their implications both within the context of the use-case but also for their broader adoption within the creative industries.

\section{Discussion}
Guiding the decisions and recommendations for the responsible innovation of this meta-story  tool were four broad themes as drawn from a mapping of the literature shown in \ref{tab1}.

\begin{table}[tb]

\caption{Themes from the Literature on the Implications of CAI for Children.}
\label{tab1}
\begin{center}
\begin{tabular}{p{2cm} p{9cm} } 
\toprule
Theme & Description\\
\midrule
1 &  Cognitive \& Linguistic Development (e.g. Educating Youth / Learning / Accessibility). \\
2 & Moral Care and Social Behaviour (e.g. Civility / Relationships / Child-Agent Interaction).  \\
3 & Ethical, Regulatory and Legal Aspects of Voice Agents for Children (e.g; Privacy / Security).\\
4  & Inclusivity (e.g. Gender / Race / Bias). \\
\bottomrule
\end{tabular}
\end{center}
\end{table}

We discuss these themes with reference to the design choices made of this meta-tool and discuss their implications. 

\subsection{Child cognitive and linguistic development}

Research shows that child-agent interactions have implications for cognitive, linguistic and educational development \cite{xu2019young,aeschlimann2020communicative}. IPAs are often described as fundamentally different to interactions between humans \cite{aeschlimann2020communicative} with vast potential for supporting children’s learning and development. Research suggests that children have a propensity not to share information, instead occupying a ‘silent world’ in which children don’t communicate about what they are seeing around them \cite{aeschlimann2020communicative}. The meta-story  tool needed to improve the communicative interaction between IPA and children or reduce it where it was harmful. What is clear in the literature is the level of children’s enjoyment of IPAs despite children not having the same expectations of IPAs as they do humans.  Instead, the technology opens up the opportunity for young children to explore knowledge, especially for those unable to read yet \cite{lovato2019hey}]. In this case, IPAs provides access to internet searching and speeds up the development of children's ‘question-asking behaviour’, something which is also explored in other aspects of the literature on semantics and invariance \cite{signorini2009if}.
Research on specific tools like Amazon’s Alexa reveals further considerations with respect to cognitive development.  Lopatovska et al., describe how children uniquely used Alexa for telling the time, perhaps because their time-telling skills are still developing, and, that unlike adults, that they did not use Alexa for games. However, more work is needed to understand the positioning of Alexa (and other IPAs) in children’s information landscape \cite[p.994]{lopatovska2018personification}. Specifically, the benefit that IPAs provide young people means that they can access information which would normally require the ability to read and write \cite{lovato2019hey}. In the design of the met-chat tool, such considerations became important.

As discussed, the relational aspects of the interaction such as ensuring fun and enjoyment, enabling engagement and `exploration' all enable children to develop functional skills. However, concerns remain about the extent to which young children are understood by a voice agent which suggests there is a need to better support children and their parents as voice agents become a greater source of answers to their questions \cite{lovato2019hey}. Importantly, the need to regulate young children’s use of voice agents is still required and a more robust approach to gathering child/parent data is required as much of the data was self-reported and there is potential for bias \cite[p.388]{lovato2019hey}. 

Further research describes likely impacts upon the cognitive development of children and outlines areas for future research on the ‘functioning of children’ \cite{biele2019might}. Special considerations are noted because of the personal and natural nature of voice communication and there are suggestions that IPA can affect linguistic habits of children, particularly with respect to politeness affecting `their 175 interpersonal dealings later in life' \cite{biele2019might}.
Finally, CAI is reported to encourage children to expect gratification or `immediate responses to their requests' \cite{wiederhold2018alexa}. Some studies suggest IPA seem more real to children and they see them as friends / companions / BFF’s \footnote{https://interestingengineering.com/research-says-kids-will-be-bffs-with-robots-in-thefuture} \cite{EngUK-robot-friends,biele2019might,pearson2014creating} - there are also concerns about reinforcing bad behavior or undesirable traits such as incivility e.g. how agents reward proper pronunciation, instead of politeness and manners \cite{druga2017hey}. Fears about the effects on social relationships - where the anthropomorphised voice agent becomes an `imaginary friend', listening to the children and harbouring their secrets are noted \cite{biele2019might}. In this regard, speech and thereby anthropomorphism can be seen to affect humanisation \cite{schroeder2016mistaking}. These aspects relate to the inclusion of children with impairments and disabilities. While the benefits for entertainment and accessibility seem clear, much research stresses the developmental aspects of how children acquire, process information and how they then might ultimately translate that into the world. These considerations formed a key part of the audio and technological development of the tool.

\subsection{Speech and linguistic development}

We found that there is much research on the way in which CAI understands childrens’ speech with a corpus of work on the analysis of language / developmental aspects \cite{monarca2020doesn} critical to the responsible design of AI Fan Along. As previously alluded to, children’s speech is not yet developed and CAI are regularly found to misunderstand and research has explored whether CAI is able to uncover language discrimination in children \cite{monarca2020doesn}. The literature suggests there is a need for inclusive solutions.
Druga et al.’s study of child-agent interaction (Alexa, Google Home, Cozmo and Julie Chatbot) \cite{druga2017hey}, provides one such example posing a series of questions to children (aged 3-10 years) related to trust and their experiences of the interaction. They found child-agent interactions were particularly revealing about children’s reflections of their own intelligence in comparison to that of the agents. The same study suggested that ‘different modalities of interaction’ may change how children perceive their own intelligence in comparison to agents. Agent voice, tone and friendliness are regularly mentioned as important considerations in ensuring interactive engagement and facilitating understanding and interactivity through expressions of characters’ ‘happy eyes’, for instance. This echoes the literature on social robots which promotes the importance of tone and voice pitch, humour and empathy. We suggest that much could be applicable to voice agents where the voice pitch is seen to have a ‘strong influence’ on user experience and enjoyment \cite{niculescu2013making}. Further, in order to better child understanding of systems, research indicates that designers ought to consider embedding into design a transparent mechanism of explaining why an agent can/cannot answer a particular question to help in re-framing it to the child, and ensuring better understanding like human interaction \cite{moreno2000engaging}. These small design considerations are important for ensuring that agents become more like companions than foes and link to issues of trust and transparency.

\subsection{Moral care and social behaviour}

Much of the CAI literature speaks to debates about moral care and social behaviour. The Human-Robot Interaction (HRI) literature relates closely to this (Ayanna Howard’s research provides clear examples) \cite{howard2018ugly} and the field for some time has looked into child-robot interaction and its effects on non verbal immediacy and childrens’ education \cite{chang2010exploring,kennedy2015higher,kennedy2016heart}, and how people treat computers, TV and New Media like real people \cite{mullen1999media}. Mayer, Sobko and Mautone’s proposed Social Agency Theory \cite{mayer2003social} argues that the social cues of a computer (e.g., modulated intonation, human-like appearance) encourage people to interpret the interaction with a computer as being social in nature. Indeed, some users report having emotional attachments to their voice agents \cite{shead2017report} and this is often debated in the literature because it infers ‘humanness’ - when some claim human-like feelings should be reserved for human interaction \cite{porra2019can}.
Research suggests that humans are more likely to engage in deep cognitive processing to make sense of what an artificial agent is saying and communicate accordingly. Children are shown to form bonds with robots and react with distress when they are mistreated \cite{mayer2003social} but associate mortality with living agents and less so robots and non living agents, which is seen to relate to them showing less moral care/ less involvement in sharing \cite{sommer2019children}. Some suggest interaction with CAI could hinder pro-social behaviour and to investigate repeated interaction over time. As such testing of the tool in this regard was suggested. A further study by Bonfert et al.’s study responds to the media’s portrayal of how children ‘adapt the consequential, imperious language style when talking to real people’ \cite[p.95]{bonfert2018if}. The experiment involved rejection when children made impolite demands, and found they adapted and behaved more outwardly politely, saying please, etc. However, many reported feelings of discontent toward the AI. Our research revealed several attempts to understand child civility with machines and spoken dialogue systems \cite{burrows1992not,potamianos1998spoken,arunachalam2001politeness}.

Finally, from a user-gender perspective, we were curious about considerations across variables. Research suggests no gender differences with respect to politeness, whereas males expressed more frustration \cite{oviatt2000talking}. As children are still learning how to formulate speech and infer meaning from interaction, it was noted that designers should accommodate and be responsive to the different languages of child users of varying ages and demographics. Collection of large scale data on children of different ages and backgrounds to pull out the ‘idiosyncratic features’ of children’s spoken word was also recommended when personalising CAI \cite{oviatt2000talking}.

\subsection{Regulatory and legal aspects of voice agents for children}

Acknowledging a recent systematic review of ethics and children-computer interaction \cite{van202018}, we find many ethical issues arising from the use of CAI, particularly with respect to surveillance \cite{mclean2019hey}, privacy and security. This results in a need for transparent design, education and regulation. For instance, studies describe IPAs as posing ‘unique problems’ concerning surveillance; i.e. they can be activated by anyone asking it questions, potentially getting access to personal information \cite{hoy2018alexa}. Research suggests that ‘major security risks’ are mitigated by voice printing systems. Children are however especially vulnerable to cyber-attacks and there are perceptions that systems are listening ‘at all times’. 

Children’s privacy is vital \cite{hoy2018alexa} (e.g. the case of surveillance and Mattel's `Aristotle'\footnote{https://www.theverge.com/2017/10/5/16430822/mattel-aristotle-ai-child-monitor-canceled}) because all interactions are recorded and analysed \cite{biele2019might}. Much of the current research debates the role an IPA ought to play with respect to safeguarding and violation of the law e.g. if a child reveals they are being abused. In order to tackle these issues, research suggests that designers ought to consult their own values \cite{biele2019might}.
Much of the research suggests a need to manage parental/ user expectations. Research suggests that children do not show awareness of the fact that the gadgets recorded interactions, whereas parents do \cite{mcreynolds2017toys}. Parents express concern about online privacy with respect to internet connected devices as well as concerns about recording and monitoring child activity and what data is held by companies \cite[p.5201]{mcreynolds2017toys}. Parents also are seen to be concerned over control and supervision, citing a lack of time to go through hundreds of recordings even if they were made available \cite{horned2020conversational}.

Conversely, it is also reported some parents find it useful to monitor their children using recordings as research suggests that parents would not wish to share their child’s recording on social media \cite{mcreynolds2017toys}. This is at odds somewhat with the findings from the children (from the same study) \cite{mcreynolds2017toys}. In this study many children did not know the device was recording and some were reported to have tricked the system through secretly wanting to speak to the device at a fair distance from their parents (2 out of 4 participants said they would tell a toy/device a secret) \cite{mcreynolds2017toys}. This highlights the need to consult both parent and child about these key issues and shaped our discussions about future qualitative work involving children and parents. Research recommends that in order to improve security and privacy: designers might 1) to include ‘visual recording indicators’ - to raise transparency and show off the capability of the device, 2) offer parents the opportunity to to engage with privacy decisions, 3) consider trust and consent - on the one hand providing the ability for parents to monitor their children might safeguard them but also poses ethical and trust issues \cite{mcreynolds2017toys}. 

Finally, research suggests that flexible interaction is important. For instance, being able to ask questions that they choose themselves to enforce existing child privacy protections through regulation \cite{mcreynolds2017toys,horned2020conversational}. Further, the same study found that children would learn quickly and develop new ways to interact with technology flexibly \cite{mcreynolds2017toys}. Van Riemsdijk et al. investigated the ethical issues surrounding creating ‘socially adaptive electronic partners’ \cite{van2015creating} and also emphasized flexibility. For instance, it was important to consider the context and how adaptive the technology is. For example, violating certain norms such as freedom, privacy etc, only if it is in the best interest of the user or the greater good, i.e. the case of an accident and releasing medical data \cite[p.1204]{van2015creating}. Flexible systems might `alleviate ethical concerns' providing `contextual integrity' \cite{nissenbaum2004privacy}. The need to ensure that systems ought to prevent unethical use, e.g a school using technology to find out if a child is skipping school is noted. Notwithstanding the limitations of contextual ethics, the importance of considering the contextual use and the everyday ethical norms which govern user behaviour remains pertinent.

\subsection{Building transparent and trustworthy CAI}

Issues of trust and transparency regularly emerge with respect to CAI ethical design \cite{mcreynolds2017toys}. Transparency has been at the forefront of the AI ethics debate as it is a tool which helps to generate trust and ultimately understanding in technology. The recent focus on transparency has led to some innovative modelling of smart assistants in order to tackle the issue \cite{geeng2020egregor}. Following our research we were clear that designers might consider explicit and implicit ways of ensuring transparency in CAI design to build respect and trust.This links to notions of fairness and inclusivity. 

Fairness is a key concept in the development of CAI technology for children. In AI, and ML field in particular, practitioners call for fairness as a solution to promote inclusivity and overcome bias (i.e., algorithmic and data bias) \cite{jain2020imperfect}. Many interesting approaches have been proposed to approach fairness in AI, such as ML AI Fairness by IBM \cite{bellamy2019ai}; and FATE: Fairness, Accountability, Transparency, and Ethics in AI toolkit \cite{birdfairlearn}. Google has also released a version of what they called Fairness Indicators \cite{xu2019fairness}, which is mainly a suite of tools that enable regular computation and visualization of ‘fairness metrics’ for ML models. In 2020 they presented ML-fairness-gym a set of components for building simple simulations to explore long-term impacts of ML models \cite{long2020ai} but many of the attempts of companies have been accused of tokenistic ethics washing.

In order to promote inclusion, much of the literature focuses on negative gender stereotypes in IPAs particularly with respect to women \cite{brahnam2012gender,danielescu2020eschewing}. Key research including UNESCO’s 2019 paper ‘I’d blush if I could’ set the scene, voicing concern about assigning gender to voice assistants and the ‘troubling repercussions’ vis a vis children’s digital skills development \cite[p.85]{Id-blush}. Additionally, much research draws attention to the issue of gender in design - rather than gender being implicit to voice - the listener assigns gender to the voice \cite{sutton2020gender}. It is suggested that until at least mid 2017, agents were evaluated as perpetuating gender stereotypes \cite{Id-blush}. There is also interesting work on misuse and abuse of social agents \cite{brahnam2012gender}. Gendered aspects of voice are not the only elements to consider: the branding, the appearance, the quality of the voice, specific pronunciations, etc are also important \cite{sutton2020gender}. In the broader literature, Pearson \& Borenstein looked into the ethics of designing companion robots for children - they suggest that an unexplored area is that of gender, which is something which has been a focus with respect to CAI in terms of persona and accent \cite{pearson2014creating}. For instance, one study found that if a robot has a male or female tone of voice, this will seriously affect the way we interact with it \cite{siegel2009persuasive}. Similarly, research found that people trust a female voice more and found it to be more persuasive \cite{crowelly2009gendered}. Coeckelbergh \cite{coeckelbergh2011humans} suggests that this is simply reflective of our daily feelings and preferences with respect to gender norms and expectations reflective of stereotypes \cite{nass2005wired} and others talk about how humans assign their own gender to robots suggestive that one should neither gender technology, nor racialise it \cite{ogunyale2018does}. Some scholars suggest that males prefer male agents and female, female agents. This has paved the way to thinking about gendering CAI e.g. \cite{donald2019societal} - who notes that the default voice for IPAs is almost always feminine and that their names are also female ‘Cortana and Alexa’ - indicative of a social signalling of gendering agents from embedded design - that their voice to language use and content. The ‘neutral’ Google Home is described as gender-less but only in name as it’s voice is female - which is the same for Siri  \cite{Fan-2021,fan2020panel}.

There is also increased focus on racial bias and injustice in technology \cite{atanasoski2019surrogate}.  Human-agent (chatbots) interaction is influenced by racial mirroring - affecting interaction with agents with respect to ‘personal interpersonal closeness, user satisfaction, disclosure comfort and desire to continue interacting’ \cite{liao2020racial}. The design implications are clear - that ‘racial mirroring facilitates the interpersonal relationship between client and agent’ \cite[p.~430]{liao2020racial}. This should be borne in mind when customising personas of (in their case) therapeutic agents, and more generally other kinds of agents\footnote{The authors note the limitations of generalising these findings beyond the setting; therapeutics and the geographical context; the US.}. Recent research describes how the white, feminine voice “reflects characteristics of white femininity in voice and cultural configuration for the purposes of white supremacy and capitalistic gain”, projecting white supremacy  \cite{moran2020racial}. Others refer less to vocal cues relating to race and instead look at content and the culturally value-laden positioning of what subjects are deemed appropriate or not \cite{schlesinger2018let}. These findings indicated to the team that in terms of the meta-story chat tool  it would be important to go beyond the voice when considering gender and racial issues in CAI design and to consider what is appropriate content for a particular use and what an appropriate response from a user would be.
This scoping provided the research team with a clear approach from which to indicate recommendations and suggestions for the design of AI Fan Along. We outline these in the following section. 

\section{Design decisions: AI Fan Along}

We now draw together the discussion points toward what resulted in design recommendations for the responsible development of the meta-story  tool. Informed by the literature and in consultation with industry, we firstly proposed a series of broad ethical considerations for  developers of a meta-story chat tool for children: 


\begin{itemize}
    \item[Q1.] What data will be collected?
    \item[Q2.] How will the collected data be used?
    \item[Q3.] How far and in relation to which regulations has the AI safeguarded children’s safety and privacy?
    \item[Q4.] How do we develop a child-friendly and engaging CAI and what behaviours should it exhibit?
    \item[Q5.] How do we reflect on and mitigate against bias?
    \item[Q6.] How do we ensure inclusive, responsible innovation and use participatory design techniques?
    \item[Q7.] What technology and approaches should be adapted to provide moral care and direct pro-social behaviour?
\end{itemize}

Using these broad questions as a base-line, we draw together the discussion to describe how we approached these with respect to (a) regulatory and legal (b) cognitive and linguistic development (c) inclusivity and (d) moral care and social behaviour as identified in the literature. 

\subsection{Ethical design of meta-story  tool AI Fan Along}

\subsubsection{Regulatory and legal aspects of CAI}

The ethical considerations of this meta-story chat tool were primarily concerned with data, privacy and user-security. 
Attending first to Q1 and data collection, we were conscious that the meta-story tool would collect voice recordings of the child-agent interaction - as a consequence, designers and developers must consider hosting and the security of the chosen system architecture. We proposed that an intelligent data privacy solution be implemented, including the gathering of consent from the parents and carers in line with  data protection and privacy - particularly important when considering third party/external industrial collaboration. Additionally, we proposed that particular attention should be given to parental permissions and levels of control. Testing with users and parents would be paramount in its further development. 

In response to Q2 about the use of the data collected, there are clear concerns about surveillance in CAI and the extent to which AI voice assistants are always listening and the efficacy of wakewords. We recommended that CAI should not run as a background process, but rather should provide parents with the control to turn it on (e.g. directly after a TV show in order to start discussion between CAI and child). Transparency is of course key to this. We therefore suggested that CAI development should be clear about what data is collected, where it will be stored, as well as acting in compliance with GDPR. Parents should be asked to provide consent for the use of personal data in the development of the technology.

With respect to Q3 about how far and in relation to which regulations has the AI safeguarded children’s safety and privacy, there is a need to examine children’s privacy, safety and security by providing identity protection, detecting harmful content and by focusing on location detection and biological/ psychological safety. UNICEF is clear that another risk for children pertains to inclusion and equitability. Ensuring that systems are checked to mitigate against historic bias which may stand in the way of children’s fair chances in life becomes a key point of ethical reflection. Research debates the role that a voice assistant ought to play with respect to safeguarding and violation of the law, for example; if a child were to reveal they are being abused. In the UK children can consent to information services at age 13 enabling them to engage freely with the internet, which is an important and largely unavoidable tool. We recommend that designers are transparent about their decision making with respect to safeguarding and do so in line with litigation and child privacy law (see OFCOM and the DCMS’s Online Harms White Paper, 2019 \cite{dcms2019harms}).

\subsubsection{Cognitive \& linguistic development}
Concerning Q4 concerning the behaviour and friendliness of agents, we proposed that designers consider their duty to consider how this impacts child development. For instance, child-friendly CAI can have a number of educational and commercial benefits and its personalisation can be very effective in engaging children in storytelling. A solution that presents CAI agents as personalised persona, based on show script scenario, allows the development of a more friendly, emotional, civil and engaging CAI. With respect to the meta-story chat tool we drew attention to three dimensions related to personalised CAI: (1) what is personalized, i.e, content, user interface, etc; (2) for whom is it personalized, i.e., sensibility of a child's context; and (3) the level of automation of personalisation. Relatedly, CAI design should consider the speaker’s variability, including age and emotion etc. This improves both the personalisation and broadens the inclusivity of CAI.

\subsubsection{Inclusivity}
As discussed inclusivity is a key consideration  relating closely to the broad prompts outlined in Q5 and Q6 related to bias and participatory design. We noted that many of the adopted practices to ensure fairness are limited to quantitative techniques, e.g., statistical models or tools that mitigating algorithmic and data biases, and assess fairness by sampling uncertainty \cite{KallusMZ20}, or de-biasing gender \cite{sun2019mitigating}. In order to ethically design CAI for children, we proposed that these methods engage with the relevant ethical literature outside of the NLP or AI fields \cite{blodgett2020language}.
In order to ensure fairness in CAI design, we called for an inclusive approach in the early stages of the design process. For example; inclusive methods to ideate answers to key questions like how to develop transparent algorithms and models that mitigate bias; e.g. adopting a task orientated dialogue system to avoid pitfalls of algorithmic bias. At all stages, we proposed that designers should consider how bias may have seeped into the development of  CAI - pertinent with respect to all aspects of CAI, not just the voice.

With respect to Q6 about inclusive design, we suggest that the design of CAI should be participatory \cite{kumar2018co,yip2019laughing,piccolo2021chatbots}. We note how children are so often not included in co-production, though research involving the views of younger people are emerging \cite{hasse2019youth}. By involving children and their parents in the design, it would be feasible to explore how far children use agents for entertainment, learning and more, especially with respect to the thematic areas we describe, particularly in the testing phase and for supporting positive child development. This was suggested for further research and development.This kind of user-involvement should keep participants as fully informed as possible about the objectives and procedures of the research to improve AI literacy \cite{long2020ai}. Indeed, deception of participants (deliberately mis-representing the purposes and aims of the study) must be avoided whenever possible and any deception should be revealed during debrief interviews with parents/guardians. 

We noted that it is not out of the question that designers may need to employ some deception during the ‘field tests’ should there be issues with the proposed prototype and/or AI voice recognition. This should be limited to obfuscating the mechanisms by which children’s interactions will be tracked, and in some instances may require responses from the prototype to be selected by researchers rather than the AI.

In advocating a participatory approach, designers must ensure that parents/ legal representatives understand consent, the objectives, any potential risks and the conditions under which the research is to be conducted. They should have been informed of the right to withdraw the child / young person from the work at any time and have a contact point where further information about the work can be obtained. 

Further, we advised that designers of CAI should consider the potential vulnerability of children to exploitation in interaction with adults (potential power relationships between adult/child) in any testing and how this might affect the child’s right to withdraw or decline in participating. We suggest that designers provide information about the task to children in an accessible way, properly explain data gathering and protection and manage expectations. We recommend that designers approach families in a timely way to ensure that children have time and opportunity to access support in their decision making about taking part. Where participants are not literate, verbal consent may be obtained and then documented. Every effort should be made to deal with consent through robust dialogue with both children and their parents. Whenever practical and appropriate, a child’s assent will be sought before including them in the research. Future research should consider error scenarios in order to consider unforeseen risks and ethical concerns \cite{arunachalam2001politeness}.

\subsubsection{Moral care and social behaviour}

Finally addressing Q7, it is pertinent to ask what technology and approaches should be adapted to provide moral care and direct pro-social behaviour. As reflected in this paper, different approaches and architectures pose distinct challenges for developing safe and responsible CAI that attend to the aspects of moral care. One key consideration is the level of freedom versus constraint that is required over NLG. For example, rule and frame-based approaches involve tightly scripted dialogues and require the designer to devise appropriate response strategies for the potential directions the dialogue may take. In retrieval-based and E2E approaches, the quality of the corpus from which responses are selected or generated is evidently important and compared to rule-based or slot-filling approaches, there is less precise control over what response is generated. With retrieval-based systems, the possible range of responses in the corpus can be checked for suitability, but it is possible that seemingly harmless responses, when produced in a different conversational context, could produce a different meaning.

As E2E systems are designed to mimic human-to-human conversations, the quality of the training data will impact on model predictions. Stringent data preprocessing efforts will be required to develop E2E systems that generate content suitable for younger audiences. Furthermore, Gehman et al. \cite{gehman2020realtoxicityprompts} demonstrate that even after implementing profanity filters on training data and fine-tuning on ‘appropriate’ data, systems can still produce toxic content. Consequently, ensuring the safety of a dialogue system requires more than removing profanities from a dataset. Harmful societal biases e.g. gender bias \cite{dinan2019queens,liu2019does} are often contained within datasets, and while Dinan et al. \cite{dinan2019queens} demonstrate that it is possible to reduce the impact of gender bias in dialogue systems, ensuring against all forms of stereotyping and representational harm in E2E systems is a complex and difficult task.

Retrieval-based and E2E approaches aim to increase the human-likeness of CAI agents, which affects how users perceive them. Moreover, some argue that CAI agents should emulate more precisely human-like behavior \cite{ahmad2018review,paikari2018framework}. In the context of child-friendly CAI, this arguably raises many ethical concerns related to trust and child protection. 

Finally, CAIs capable of engaging conversation, designed to utilise relational strategies may influence the child’s perception on the humanness of the agent and influence their behaviour \cite{lovato2019hey}. We also highlight the importance of these CAI agents to identify themselves as bots and to provide specific answers and clarify it to the user when the context/question is not comprehensible.

\section{Conclusion}

The development of CAI in the creative industry for children has been limited and there is a growing need to connect theory and practice. Indeed, much of the research has been about the impact on children, as opposed to with and for \cite{hodge2017restricted}. The field in its current manifestation presents, at best, an inconsistent approach to the systems explored here, often with a need to join up the ramifications of situating such technologies within the home with the implications for children. As momentum grows in the overall ethics of AI, the inherent biases and assumptions underpinning the technical methodologies require the utmost scrutiny when applied to vulnerable groups such as children. This pilot case-study highlights the unique concerns located within AI storytelling tools for children. The reflections of the design choices made and recommendations provide a starting point from which to extrapolate and build on the field of AI ethics for children. However, further research to provide greater depth and richness of perspectives is recommended and significant remedial work is required at all levels of the design process across stakeholders inclusive of developers, content makers, users (including parents and guardians from all backgrounds) and importantly, educators and regulators.

\section*{Acknowledgements}
This work was funded by the XR Stories: Young XR grant, AI Fan Along, and the Digital Creativity Labs, jointly funded by EPSRC/AHRC/Innovate UK, EP/M023265/1 and the Humanities and Social Change International Foundation.

We would also like to thank our industry partners.
\bibliographystyle{unsrtnat}
\bibliography{CAIChildren}

\end{document}